\documentclass[12pt]{article}
\newcommand{\beq}{\begin{equation}}
\newcommand{\eeq}{\end{equation}}
\newcommand{\bra}{\begin{array}}
\newcommand{\era}{\end{array}}

\newcommand{\al}{\alpha}

\newcommand{\de}{\delta}

\newcommand{\ep}{\epsilon}
\usepackage{latexsym}

\begin{document}
\centerline{\Large\bf A Short Note on Multi-bion Solutions}

\vspace{2cm}
\centerline{Rajsekhar Bhattacharyya\footnote{E-address: rbhattac@sun.ac.za}}
\vspace{1cm}
\centerline{Stellenbosch Institute for Advanced Study}
\centerline{Stellenbosch, South Africa }


\vspace{4.5cm}

\begin{abstract}

Here we try to construct a form of multi-bion solution in the dual description of $D3 \bot D1$-system which connects the two separated bions each made up of 2 $D1-branes$ at large distance with a single $D3$-brane of four unit of magnetic charge at origin. Further we interested in the soluions which can interpolate between arbitrarily separated bions and single $D3$-brane with arbitrary amount of magnetic charges and we find that it is probably not possible to have the solution in each case.

\vspace{0.5cm}


\vspace{0.5cm}
{PACS number(s): 11.25.Uv}
\end{abstract}



\newpage
The study of $D$-brane has brought significant changes for the understanding of string theory particularly for the study of the geometry of the space-time, for reviews see [1,2,3,4]. Much of the progress has come about by studying the low energy dynamics of the $D$-brane world volume which is given by the Born-Infeld action. This world volume theory on $D$-branes has many fascinating features. There are possibilities in a $Dp$-brane theory that through the appropriate exitation of fields, we can get objects which may be interpreted as $Dp'$-branes of lower or higher dimensionality. In this context it is worth to mention that study of these systems where $D$-strings ending in $D3$, $D5$ and $D7$-branes becomes more important as these systems give rise to non-commutative geometry through non-commutative fuzzy funnels [5,6,7] and for recent interesting developments in this direction can be found in [8,9,10]. In particular in $D3 \bot D1$ system we find the geometry of fuzzy two sphere.

\vspace*{1cm}
For $D3 \bot D1$ system in [5] the nonabelian Born-Infeld action is taken for $N$ $D1$-branes in a flat background and it is shown there that the transverse coordinates can be interpreted as a collection of $D1$-branes attached to $D3$-brane. This is a dual description of magnetic charge in $D3$ brane with abelian Born-Infeld action and it can be shown that for large $N$ both the descriptions predicts the same geometry of the fuzzy funnel with same energy configuration. The $D3 \bot D1$ system is supersymmetric, which is clear in both pitures, since the solutions describing the intersection are BPS in nature. In the case of nonabelian Born-Infeld action for the $D1$-branes the BPS equation becomes the Nahm equation.

\vspace*{1cm}


First we review the $D3\bot D1$ brane system. 
We suggest that D3-brane has more than one scalar describing
transverse fluctuations. we denote the world volume coordinates by
$\sigma^a$, $a=0,1,2,3$, and the transverse directions by the
scalars $\phi^i$, $i=4,...,9$. In D3-brane theory construction,
the low energy dynamics of a single D3-brane is described by the
Born-Infeld action in static gauge
$$
S_{BI}=\int L=-T_3 \int d^4\sigma\Big(-det(\eta_{ab}+\lambda^2\partial_a \phi^i \partial_b \phi^i +\lambda F_{ab})\Big)^{\frac{1}{2}}
$$
where $F_{ab}$ is the field strength of the $U(1)$ gauge field on
the brane and $\lambda =2\pi\alpha'=2\pi\ell_s^2 $ ($\ell_s$ is
the string length). By exciting one scalar field and calling it $\phi$ and magnetic field $B^i=\epsilon^{ijk}F_{jk}$ and setting zero values to the other fields, the energy is evaluated for the
fluctuations. For static configuration
$$\bra{ll}
E=T_3 \int d^3\sigma\Big ( 1 +\lambda^2 ( \mid \nabla\phi
\mid^2 +B^2 )+ \lambda^4
((B.\nabla\phi)^2 \Big ) ^{\frac{1}{2}} .
\era$$ 
$$= T_3 \int d^3\sigma\Big( \lambda^2 \mid \nabla\phi +\stackrel{\rightarrow}{B} \mid^2 +(1-\lambda^2 \nabla\phi.\stackrel{\rightarrow}{B})^2 \Big)^{\frac{1}{2}} $$

\vspace*{1cm}
So requring minimum energy condition we find
$$\nabla\phi = \pm\stackrel{\rightarrow}{B}\eqno(1)$$
which coinsides with the BPS condition for the magnetic monopole [11,12,13] and for that we find minimum energy $E_0$ to be
$$E_0 = T_3 \int d^3\sigma\Big(1-\lambda^2 \nabla\phi.\stackrel{\rightarrow}{B})\eqno(2) $$
Now using Bianchi identity the simplest solution coresponding to the bion spike and is given by
$$ \phi = \frac{N}{r}~~~;~~~\stackrel{\rightarrow}{B}=\mp \frac{N\hat{r}}{r^2}\eqno(3)$$
where $r^2= (\sigma^1)^2 + (\sigma^2)^2 + (\sigma^3)^2$ and $N$ is an integer due to quantisation. This magnetic bion solution correspons to $N$ superposed $D1$-branes attached to the $D3$-brane at the origin. As we have have identified the monopoles as BPS object so there is no force between two such objects when they are kept at finite distances. Thus we can have the general solution as the sum of $k$ such spikes which can be given as  
$$\phi = \sum^k_{a=1}\frac{N_a}{\mid \stackrel{\rightarrow}{r}-\stackrel{\rightarrow}{r_a}\mid}\eqno(4)$$ where $N_a$ charge is at radius vector $\stackrel{\rightarrow}{r_a}$.

\vspace*{1cm}
Now we consider the dual description of the $D3\bot D1$-system in D-string theory. The low energy dynamics of $N$ D-strings in a flat background can be described by the non abelian Born-Infeld action 
$$S=-T_1\int d^2\sigma STr \Big( -det(\eta_{ab}+\lambda^2 \partial_a \phi^i Q_{ij}^{-1}\partial_b \phi^j )det Q^{ij}\Big)^{1\over2}$$
where $ Q_{ij}=\de_{ij} +i\lambda \lbrack \phi_i , \phi_j \rbrack$. Again we here assume static gauge where the two worldsheet coordinates are idntified with $\tau=x^0$ and $\sigma=x^9$ and the transverse scalars $\phi^i$'s are $N\times N$ matrices transforming in the adjoint representation of $U(N)$ gauge symmetry. Now combining two determinants into single one 
the action can be rewritten as
$$S=-T_1\int d^2\sigma STr \Big( -det\pmatrix{\eta_{ab}&
\lambda \partial_a \phi^j \cr
-\lambda \partial_b \phi^i & Q^{ij}\cr}\Big)^{1\over2}.$$ which gives energy as
$$E=T_1\int d\sigma STr \Big( \lambda^2(\partial_{\sigma}\phi^i \pm \frac{i}{2}\epsilon^{ijk}\lbrack \phi_j , \phi_k \rbrack)^2 +(1\mp\frac{i}{2}\lambda^2 \epsilon^{ijk} \partial_{\sigma}\phi^i \lbrack \phi_j , \phi_k \rbrack)^2 \Big)^{1\over2}$$
for minimum energy condition we find the solution
$$\partial_{\sigma}\phi^i \pm \frac{i}{2}\epsilon^{ijk}\lbrack \phi_j , \phi_k \rbrack =0\eqno(5)$$ which can be identified as the Nahm equation. This gives the minimum energy as 
$$E'_0= T_1\int d\sigma \Big(N\mp \frac{i}{3}\lambda^2 \epsilon^{ijk} \partial_{\sigma}Tr(\phi_i  \phi_j  \phi_k ) \Big)\eqno(6)$$

\vspace*{1cm}
The solution of the equation of motion of the scalar fields $\phi_i$, $i=1,2,3$ represent the D-string expanding into a D3-brane analogous to the bion solution of the D3-brane theory [5]. The solutions are 
$$\bra{lc}\phi_i =2R(\sigma)\alpha^i ~~~;~~~\lbrack \al^i , \al^j \rbrack=i\ep^{ijk}\al_k \era$$
where $R(\sigma)=\frac{1}{2\sigma}$ and $\alpha^i$'s are $N\times N$  dimensional generators of $SU(2)$. 
Now for these $\alpha^i$'s we find $\sum_i(\alpha^i)^2=CI_N$ where $I_N$ is the $N\times N$ identity matrix and $C$ is the Casimir. For $C=N^2-1$ we find
$$R(\sigma)=\lambda\sqrt{\frac{Tr[\phi^i(\sigma)^2]}{N}}\approx \frac{N\pi \l^2_s}{\sigma}\eqno(7)$$ 
which is the physical radius of a fuzzy 2-sphere. This geometry can be compared to the $D3$-brane solution (3) after identifying $\sigma \rightarrow \lambda\phi$ and $R \rightarrow r$ together with the energies given by
(2) and (6) at large $N$.

\vspace*{1cm}
Now it can be investigated that how one can obtain an analogue of the multi-bion solution (as given in (4)) in the dual description of $D3\bot D1$ system. In other words we can seek for a solution which can interpolate between the boundary conditions for $\phi^i$'s for $\sigma \rightarrow \infty$ and for $\sigma \rightarrow 0$ i.e 
$$\phi^{i}(\sigma\rightarrow\infty)=diag(x^i_{1}I_{q(1)\times q(1)} + 2R\alpha^i_{q(1)},....,x^i_{k}I_{q(k)\times q(k)} + 2R\alpha^i_{q(k)})\eqno (8)$$ and $$\phi^{i}(\sigma\rightarrow 0)=2R\alpha^i_N\eqno (9)$$
where $\alpha^i_{q(a)}$ are $q(a)$-dimensional generators of an irreducible representation of $SU(2)$. The solution for $\sigma \rightarrow \infty$ represents number of parallel $D1$-brane bunches, separated in space, where for $\sigma \rightarrow 0$ we have an irreducible representation for $SU(2)$ i.e there is a single $D3$-brane. Now in [14] it has been shown that there is a solution which interpolates between (8) and (9) for $k=2$, $q(1)=q(2)=2$ and $N=4$


\vspace*{1cm}
Here we try to construct a form of such a solution which interpolates between (8) and (9). We show that such a form exists for $\sigma\rightarrow 0$ and $\sigma\rightarrow\infty$ for $k=2$, $q(1)=q(2)=2$ and $N=4$ and we also suggest for the solutions for $k=2$, $q(1)=q(2)=n$ and $N=2n$. Further as we are studying two dual theory it is very much expected to have the solutions for arbitrary $k$, $q(a)$ and $x^i_a$'s. Here investigating that fact we find that it is probably not possible to have the solution for every $k$, $q(a)$ and $x^i_a$'s.


\vspace*{1cm}

So to find the form of the solution for $k=2$, $q(1)=q(2)=2$ and $N=4$ let us assume that $\phi^{1}(\sigma)$, $\phi^{2}(\sigma)$ and $\phi^{3}(\sigma)$ are given as

$$\phi^1(\sigma)=2\left(\matrix{-A(\sigma) &0 &0 &0 \cr 0 &-B(\sigma) &0 &0\cr 0 &0 &B(\sigma) &0\cr0 &0 &0 &A(\sigma)}\right)$$
$$\phi^2(\sigma)=2\left(\matrix{0 &C(\sigma) &0 &0 \cr C(\sigma) &0 &D(\sigma) &0\cr 0 &D(\sigma) &0 &C(\sigma)\cr0 &0 &C(\sigma) &0}\right)\eqno (10)$$
$$\phi^3(\sigma)=2\left(\matrix{0 &iC(\sigma) &0 &0 \cr -iC(\sigma) &0 &iD(\sigma) &0\cr 0 &-iD(\sigma) &0 &iC(\sigma)\cr0 &0 &-iC(\sigma) &0}\right)$$

\vspace*{0.5cm}
and from Nahm equation (5) we find
$$\frac{dA}{d\sigma}=-2C^2~~~;~~~
\frac{dB}{d\sigma}=2C^2-2D^2~~~;~~~
\frac{dC}{d\sigma}=2(B-A)C~~~;~~~
\frac{dD}{d\sigma}=-2BD\eqno (11)$$

and for such a system we have the boundary conditions 
$$A=\Delta +R(\sigma)~~~ ,~~~B=\Delta -R(\sigma)~~~,~~~ C=R(\sigma)~~~,~~~ D=0\eqno (12)$$
for $\sigma\rightarrow \infty$, and for $\sigma \rightarrow 0$
$$A=3R(\sigma)~~~,~~~B=R(\sigma)~~~,~~~C=\sqrt3 R(\sigma)~~~,~~~D=2R(\sigma)\eqno (13)$$

\vspace*{1cm} 
We assume 
$$A=\Delta(1-e^{-\sigma})+(1+2g(\sigma)e^{-\sigma})R(\sigma)\eqno (14)$$
and using (11) we can construct $B$, $C$, and $D$ and lastly we will get a differential equation of $g$ of order 4. Clearly from the solution of the last equation and for $\sigma\rightarrow\ 0$ and $\sigma\rightarrow\infty$ if we can find $g\rightarrow a$ where $a$ is a nonzero constant and $g\rightarrow $ finite respectively then adjusting $g$ we can find desirable $A$ which will interpolate between (12) and (13) and construction ensures that the other $B$, $C$, and $D$ also interpolates between (12) and (13) for $\sigma\rightarrow\ 0$ and $\sigma\rightarrow\infty$. Here we will show $g$ satisfy differential equtaion for $\sigma\rightarrow\ 0$ and $\sigma\rightarrow\infty$ with desirable boundary conditions.

\vspace*{1cm}
Let $G=g(\sigma)e^{-\sigma}R(\sigma)$. So from the assumed form of $A$ we find
$$-4BD^2=2C\frac{dC}{d\sigma} -\frac{1}{8C^2}\frac{dK}{d\sigma}+ \frac{K}{4C^3}\frac{dC}{d\sigma}+ \frac{d}{d\sigma}\Big(\frac{B}{C}\frac{dC}{d\sigma}\Big)-\frac{d}{d\sigma}\Big(\frac{A}{C}\frac{dC}{d\sigma}\Big)-\frac{1}{2}\frac{d^2A}{d\sigma^2}\eqno (15)$$
where
$$K=-\Delta e^{-\sigma}-\frac{d^3R}{d\sigma^3}-2\frac{d^3G}{d\sigma^3}$$
Now for $\sigma$ very near to zero we can replace $A$, $B$, $C$, and $D$ by (13) and the last equation implies
$\frac{dK}{d\sigma}=-\frac{2K}{\sigma}+ \frac{3}{\sigma^5}$
which can be approximated to $\frac{d^3G}{d\sigma^3}=-\frac{9}{4\sigma^4}$ which yields
$$G=\frac{a}{\sigma}+c_3\sigma^2 +c_2\sigma +c_1= \frac{g}{2\sigma}$$ which implies that for $\sigma\rightarrow\ 0$, $g\rightarrow 2a$. So we can modify the old $A$ by
$$A=\Delta(1-e^{-\sigma})+(1+\frac{g}{a}(\sigma)e^{-\sigma})R(\sigma)\eqno(16)$$
together with $B$, $C$, and $D$ which can be constructed from (11) and they will be the solution of the system near zero. Furher for large $\sigma$ again replacing $A$, $B$, $C$, and $D$ by (12) and from the differenial equation we find $\frac{dK}{d\sigma}=-\frac{2K}{\sigma}- \frac{3}{\sigma^5}$ which can be again approximated to $\frac{d^3G}{d\sigma^3}=\frac{3}{2\sigma^4}$ which gives $G$ goes like $\frac{1}{\sigma}$ so $g(\sigma)$ is finite. Thus we can find the solution also for large $\sigma$.

\vspace*{1cm}

Now we try to construct a solution which can interpolate between (8) and (9) for $k=2$, $q(1)=q(2)=n$ and $N=2n$.
For that we can choose $\phi^{1}(\sigma)$ as 
$$\phi^1(\sigma)=2diag(-A_1, -A_2,....,-A_n, A_n,....,A_2, A_1,)$$
for $\phi^{2}(\sigma)$ and $\phi^{3}(\sigma)$ take 
$$M_1=2superdiag(C_1, C_2,...., C_{n-1}, C_n, C_{n-1},....,C_2, C_1,)$$
$$M_2=2subdiag(C_1, C_2,...., C_{n-1}, C_n, C_{n-1},....,C_2, C_1,)$$
and we can assume 
$$\phi^{2}(\sigma)=2(M_1 + M_2)$$
$$\phi^{3}(\sigma)=2i(M_1 - M_2)$$
Now analogous to (11) we have
$$\frac{dA_1}{d\sigma}=-2C_1^2~~~;~~~
\frac{dA_k}{d\sigma}=2C_{k-1}^2-2C_k^2$$ for $k=2\rightarrow n$ and also
$$\frac{dC_n}{d\sigma}=-2A_n C_n~~~,~~~\frac{dC_k}{d\sigma}=(A_{k+1}-A_k)C_k$$ for $k=1\rightarrow n-1$
and analogous to and (13) for $\sigma \rightarrow 0$
$$A_k=(2n-2k+1)R(\sigma)~~~,~~~C_{k+1}=\frac{2}{\sqrt2}\sqrt{\frac{(k+1)(2n-1-k)}{2}}R(\sigma)$$
for $k=1\rightarrow n$ and for $k=0\rightarrow n-1$ respectively, futher analogous to (12) for $\sigma \rightarrow \infty$
$$A_k=\Delta + (n-2k+1)R(\sigma)~~~,~~~C_{k+1}=\frac{2}{\sqrt2}\sqrt{\frac{(k+1)(n-1-k)}{2}}R(\sigma)$$
for $k=1\rightarrow n$ and for $k=0\rightarrow n-1$ respectively. Now assuming the form of $A_1$ as

$$A_1= \Delta(1-e^{-\sigma})+(n-1+ng(\sigma)e^{-\sigma})R(\sigma)\eqno(17)$$
we can construct all $A_i$'s and $C_i$'s for $i=1\rightarrow n$ with one differential equation of order $2n$ of $g$ (like (15) here), where $g$ has to obey similar boundary conditions. Again if $g$ satisfy the $2n$ order differential equation with desired boundary condition we can get the solution.  


\vspace*{1cm}
Now we wish to investigate the existance of the solution which can be interpolated between (8) and (9) for arbitray $k$, $q(a)$ and $x^i_a$. For this we start to examine how $N=5$ unit magnetic charge at the origin of $D3$-brane can be separated into three bunch of bions at large distance where two of them are made of 2 $D1$-branes and one is a single $D1$-brane. So taking their separations under consideration we can seek for a solution which interpolates between
$$\phi^{i}(\sigma\rightarrow\infty)=diag(-2\Delta I_{q(1)\times q(1)} + 2R\alpha^i_{q(1)},  2\Lambda I_{q(2)\times q(2)}+ 2R\alpha^i_{q(2)},  2\Delta I_{q(3)\times q(3)} + 2R\alpha^i_{q(3)})\eqno (18)$$ and $$\phi^{i}(\sigma\rightarrow 0)=2R\alpha^i_N\eqno (19)$$ where $q(1)=q(3)=2$, $q(2)=1$ and $N=5$ as for $N=4$, $k=2$, $q(1)=q(2)=2$ the existance of the solution has been proved i.e. for four unit of magnetic charge we have a parallel pair of 2 $D1$-branes separated by $4\Delta$ for $\sigma \rightarrow \infty$

\vspace*{1cm}
Now considering the symmetry of the problem we can choose an ansatz for $\phi(\sigma)$'s as
$$\phi^1(\sigma)=2\left(\matrix{-A(\sigma) &F_1(\sigma) &F_2(\sigma) &F_3(\sigma) &F_4(\sigma)\cr F_1(\sigma) &-B(\sigma) &F_5(\sigma) &F_6(\sigma) &F_7(\sigma)\cr F_2(\sigma) &F_5(\sigma) &C(\sigma) &F_5(\sigma) &F_2(\sigma)\cr F_3(\sigma) &F_6(\sigma) &F_5(\sigma) &B(\sigma) &F_1(\sigma)\cr F_4(\sigma) &F_7(\sigma) &F_2(\sigma) &F_1(\sigma) &A(\sigma)}\right)$$

$$\phi^2(\sigma)=2\left(\matrix{X_1(\sigma) &D(\sigma) &F(\sigma) &G(\sigma) &H(\sigma)\cr D(\sigma) &X_2(\sigma) &E(\sigma) &J(\sigma) &I(\sigma)\cr F(\sigma) &E(\sigma) &X_3(\sigma) &E(\sigma) &F(\sigma)\cr G(\sigma) &J(\sigma) &E(\sigma) &X_4(\sigma) &D(\sigma)\cr H(\sigma) &I(\sigma) &F(\sigma) &D(\sigma) &X_5(\sigma)}\right)\eqno (20)$$

$$\phi^3(\sigma)=2\left(\matrix{X'_1(\sigma) &iD'(\sigma) &iF'(\sigma) &iG'(\sigma) &iH'(\sigma)\cr -iD'(\sigma) &X'_2(\sigma) &iE'(\sigma) &iJ'(\sigma) &iI'(\sigma)\cr -iF'(\sigma) &-iE'(\sigma) &X'_3(\sigma) &iE'(\sigma) &iF'(\sigma)\cr -iG'(\sigma) &-iJ'(\sigma) &-iE'(\sigma) &X'_4(\sigma) &iD'(\sigma)\cr -iH'(\sigma) &-iI'(\sigma) &-iF'(\sigma) &-iD'(\sigma) &X'_5(\sigma)}\right)$$

\vspace*{1cm}
Now as $\phi^i$'s should satisfy (18) and (19) for $\sigma \rightarrow \infty$ and for $\sigma \rightarrow 0$ respectively this implies for $\sigma \rightarrow 0$, $C \rightarrow 0$. Now from Nahms equation we find $$\frac{dC}{d\sigma}=0$$ which gives $C=0$. But if we claim that $\phi^1(\sigma \rightarrow \infty)$ satisfy (18) then  we always have third diagonal element of $\phi^1(\sigma \rightarrow \infty)$ to be non zero. So we find contradiction. So we can not have solution of the form in (20) which can satisfy (18) and (19).

Further we can choose a more general ansatz for $\phi(\sigma)$'s as
$$\phi^1(\sigma)=2\left(\matrix{A'(\sigma) &F_1(\sigma) &F_2(\sigma) &F_3(\sigma) &F_4(\sigma)\cr F_1(\sigma) &B'(\sigma) &F_5(\sigma) &F_6(\sigma) &F_7(\sigma)\cr F_2(\sigma) &F_5(\sigma) &C(\sigma) &F_8(\sigma) &F_9(\sigma)\cr F_3(\sigma) &F_6(\sigma) &F_8(\sigma) &B(\sigma) &F_{10}(\sigma)\cr F_4(\sigma) &F_7(\sigma) &F_9(\sigma) &F_{10}(\sigma) &A(\sigma)}\right)$$

$$\phi^2(\sigma)=2\left(\matrix{X_1(\sigma) &D(\sigma) &F(\sigma) &G(\sigma) &H(\sigma)\cr D(\sigma) &X_2(\sigma) &E(\sigma) &J(\sigma) &I(\sigma)\cr F(\sigma) &E(\sigma) &X_3(\sigma) &K(\sigma) &L(\sigma)\cr G(\sigma) &J(\sigma) &K(\sigma) &X_4(\sigma) &M(\sigma)\cr H(\sigma) &I(\sigma) &L(\sigma) &M(\sigma) &X_5(\sigma)}\right)\eqno (21)$$

$$\phi^3(\sigma)=2\left(\matrix{X'_1(\sigma) &iD'(\sigma) &iF'(\sigma) &iG'(\sigma) &iH'(\sigma)\cr -iD'(\sigma) &X'_2(\sigma) &iE'(\sigma) &iJ'(\sigma) &iI'(\sigma)\cr -iF'(\sigma) &-iE'(\sigma) &X'_3(\sigma) &iK'(\sigma) &iL'(\sigma)\cr -iG'(\sigma) &-iJ'(\sigma) &-iK'(\sigma) &X'_4(\sigma) &iM'(\sigma)\cr -iH'(\sigma) &-iI'(\sigma) &-iL'(\sigma) &-iM'(\sigma) &X'_5(\sigma)}\right)$$

\vspace*{1cm}
Now as $\phi^3$ should satisfy (18) and (19) for $\sigma \rightarrow \infty$ and for $\sigma \rightarrow 0$ respectively this implies for $\sigma \rightarrow 0$, $X'_3 \rightarrow 0$ where $X'_3$ is real. Now from Nahm equation we find $$\frac{dX'_3}{d\sigma}=\frac{i}{2}\Big([\phi^1,\phi^2]\Big)_{33}$$ where LHS of this equation is real while RHS is imaginary which gives $X'_3=0$. But if we claim that $\phi^3(\sigma \rightarrow \infty)$ satisfy (1) then  we always have third diagonal element of $\phi^3(\sigma \rightarrow \infty)$ to be non zero. So if we take the ansatz for the solution in the form of (21) we find contradiction. Thus it is probably not possible to find the solution which can be interpolated between (8) and (9) for arbitray $k$, $q(a)$ and $x^i_a$. 

\vspace*{1cm}

Further the whole above discussion is true when $N=2n+1$ amount of $U(1)$ magnetic charge in $D3$-brane can be separated into three bions such that two of them are bunch of $n D1$-branes each and third one is a single $D1$-brane.

\vspace*{1cm}
 So concludingly here we can predict a form of the solution which can connect a fuzzy sphere with $N=4$ to two fuzzy sphere with $N=2$. The technique which has been adopted for the solution can be extended to the cases of higher even $N$ in a straightforward way. Further we have observed that it is probably not possible to find the solution which can be interpolated between (8) and (9) for arbitray $k$, $q(a)$ and $x^i_a$. In this context we can mention that there are more general boundary conditions for solving the Nahm equation [15,16,17,18], where we can study the similar situations. But as it is well known fact that the dual description of $D3 \bot D1$ system gives the same geometry of $D3 \bot D1$ system at large $N$ i.e from the dual description we can find the fuzzy sphere as $D3$ brane with semi infinite bions at large $N$, so we may not expect analogue of (4) in the dual description for every $N$.

\vspace*{1cm}

{\bf Acknowledgment}

\vspace*{0.5cm}

I want to thank Robert de Mello Koch and Joanna L Karczmarek for helpful discussion.

\end{document}